\documentclass[aip,jcp,preprint]{revtex4-1}

\usepackage[utf8]{inputenc}
\usepackage[english]{babel}
\usepackage{amsmath}
\usepackage{mathtools}
\usepackage{amsfonts}
\usepackage{amssymb}
\usepackage{amsthm}
\usepackage{algorithmicx}
\usepackage{algpseudocode}
\usepackage{graphicx}
\usepackage{hyperref}
\usepackage{silence}
\usepackage{float}
\usepackage{geometry}
\usepackage{enumitem}    
\usepackage{tikz}
\usepackage{caption}
\usepackage{comment}
\usepackage{subcaption}

\usepackage{algorithm}
\hypersetup{
    colorlinks=true,
    linkcolor=blue,
    filecolor=magenta,
    urlcolor=cyan,
    citecolor=blue
}

\makeatletter
\renewcommand{\p@subsection}{} 
\makeatother

\begin{document}

\title{Accelerating Correlated Wave Function Calculations with Hierarchical Matrix Compression of the Two-Electron Integrals}

\author{Hongji Gao}
\affiliation{Institute for Advanced Computational Science, Stony Brook University, Stony Brook, NY, USA 11794}
\affiliation{Department of Applied Mathematics and Statistics, Stony Brook University, Stony Brook, NY, USA 11794}

\author{Xiangmin Jiao}
\email[Correspondence email address: ]{xiangmin.jiao@stonybrook.edu}
\affiliation{Institute for Advanced Computational Science, Stony Brook University, Stony Brook, NY, USA 11794}
\affiliation{Department of Applied Mathematics and Statistics, Stony Brook University, Stony Brook, NY, USA 11794}

\author{Benjamin G. Levine}
\email[Correspondence email address: ]{ben.levine@stonybrook.edu}
\affiliation{Institute for Advanced Computational Science, Stony Brook University, Stony Brook, NY, USA 11794}
\affiliation{Department of Chemistry, Stony Brook University, Stony Brook, NY, USA 11794}

\begin{abstract}
Leveraging matrix sparsity has proven a fruitful strategy for accelerating quantum chemical calculations.  Here we present the hierarchical SOS-MP2 algorithm, which uses hierarchical matrix ($\mathcal{H}^{2}$) compression of the electron repulsion integral (ERI) tensor to reduce both time and space complexity.  This approach is based on the atomic orbital Laplace transform MP2 calculations, leveraging the data-sparsity of the ERI tensor and the element-wise sparsity of the energy-weighted density matrices.  The $\mathcal{H}^{2}$ representation approximates the ERI tensor in a block low-rank form, taking advantage of the inherent low-rank nature of the repulsion integrals between distant sets of atoms.  The resulting algorithm enables the calculation of the Coulomb-like term of the MP2 energy with a theoretical time complexity of $\mathcal{O}(N^{2}\log N)$ and a space complexity of $\mathcal{O}(N^{2}\log N)$, where $N$ denotes the number of basis functions.  Numerical tests show asymptotic time and space complexities better than $\mathcal{O}(N^{2})$ for both linear alkanes and three-dimensional water clusters.
\end{abstract}

\maketitle

\section{Introduction}
\label{sec:Introduction}
A variety of computational strategies have been developed to improve computational scaling and to leverage modern parallel computer hardware.\cite{calvin2021chemrev} The compression of the electron repulsion integral (ERI) tensor has proven an effective strategy for accelerating correlated wave function calculations. Density fitting\cite{whitten1973coulombic,baerends1973hfs,vahtras1993integral,bernholdt1998fitting,hattig2005optimization,aquilante2008accurate,qin2023interpolative} (DF) and Cholesky factorization\cite{beebe1977simplifications,weigend1998ri,koch2003sanchez,epifanovsky2013general,peng2017eri,zhang2021toward} are powerful approaches to construct low-rank approximations to the ERI tensor, now widely used in practice. The continuous fast multipole methods (CFMM)\cite{white1994continuous,white1996linear,strain1996achieving,lazarski2015density,lazarski2016density} and the clustered low-rank tensor (CLR) format\cite{lewis2016clustered} have also been employed to compress the ERI tensor efficiently. Tensor hypercontraction (THC)\cite{hohenstein2012tensor,hohenstein2012communication,song2016atomic,song2017atomic,lee2021even} further exploits a four-dimensional compression strategy to achieve additional scaling advantages.

An alternative means of exploiting the sparsity of the ERI tensor is to devise an orbital basis that capitalizes on the short-range nature of electron-electron interactions. This approach underlies the projected atomic orbital (PAO), pair natural orbital (PNO), and related local correlation methods, in which the ERI tensor is represented in an orbital basis carefully chosen to preserve sparsity.\cite{pulay1983localizability,saebo1993local,schutz1999low,neese2009pno,liakos2015possible,wang2023sparsity} As a result, PNO-based approaches are particularly well-suited to local MP2 and other localized correlation methods,\cite{werner2003fast,frank2017pno} achieving essentially linear scaling for sufficiently large systems. The rapid development of efficient implementations of such approaches has been facilitated by new general-purpose computational tools for working with sparse maps.\cite{pinski2015sparsemaps,riplinger2016sparsemaps}

In the context of Hartree--Fock theory, Chow and co-workers introduced an intriguing ERI compression strategy based on a blockwise low-rank representation known as the $\mathcal{H}^2$-matrix.\cite{xing2020linear} The $\mathcal{H}^2$-matrix representation is a hierarchical block low-rank structure that enables both the storage requirements and the computational costs associated with matrix-vector multiplications to scale linearly with the matrix dimension.\cite{hackbusch2002data} The ERI tensor, $(\mu \nu | \lambda \sigma)$, describes the Coulomb interaction between products of basis functions. Defined as
\begin{equation}
(\mu\nu | \lambda\sigma) = \iint \phi_{\mu}(r_{1}) \phi_{\nu}(r_{1}) \frac{1}{|r_{1} - r_{2}|} \phi_{\lambda}(r_{2}) \phi_{\sigma}(r_{2}) \, dr_{1} \, dr_{2},
\label{eq:ERI}
\end{equation}
the number of non-negligible entries scales as $\mathcal{O}(N_{\text{bf}}^2)$ in large molecules, although the prefactor may be large.  The $\mathcal{H}^2$-matrix format achieves efficient compression by taking advantage of the diagonal dominance of the ERI and the fact that interactions between well-separated basis function pairs exhibit low-rank behavior. In fact, hierarchical matrices can be thought of as an algebraic analogue of the fast multipole method,\cite{white1996linear} and from a physical perspective, both methods leverage the same inherent data-sparsity of the ERI.  The $\mathcal{H}^2$-matrix representation has been applied to Hartree--Fock and Kohn--Sham density functional calculations, demonstrating excellent time and space complexity for large systems.\cite{xing2020fast,xing2020interpolative,xing2020linear}
The CLR approximation of Valeev and coworkers is similar in spirit.\cite{lewis2016clustered}  When applied to compute the exchange terms in Hartree--Fock calculations, CLR has been shown to provide excellent scaling, even for relatively small systems.
Beyond the ERI tensor, our groups recently demonstrated a promising method for hierarchical matrix compression of the full configuration interaction wave function.\cite{berard2024efficient}

To date, the $\mathcal{H}^2$-matrix representation of the ERI has been applied only within mean-field theories. Yet, given the opportunities for reduced scaling and improved parallelization, it appears extremely well-suited for extension to correlated wave function methods. In this work, we focus on second-order Møller--Plesset perturbation theory (MP2) as an initial target. MP2 has long been a mainstay of computational quantum chemistry, and remains a robust and efficient method for describing weak electron correlation.\cite{moller1934note,cremer2011} Recent years have seen renewed interest in MP2 with the growing popularity of double-hybrid density functionals.\cite{angyan2005van,grimme2006,chai2009,sharkas2011,goerigk2014double,martin2020} Despite its lower computational demands relative to more advanced methods such as coupled-cluster theory, the conventional implementation of MP2 still requires significant computational resources for large molecules. The time and space complexity of a naive implementation scale as $\mathcal{O}(N^{5})$ and $\mathcal{O}(N^{4})$, respectively, where $N$ is the number of basis functions. The bottleneck of the atomic orbital MP2 calculation lies in the storage and manipulation of the ERI tensor and the density matrices.  Therefore, compression of the ERI tensor is central to reducing the overall time and space complexity of MP2 calculations.

In this work, we introduce a hierarchical spin opposite-scaled (SOS-) MP2 algorithm that applies the hierarchical matrix representation to the ERI tensor within the SOS-MP2 framework. The SOS-MP2 method approximates the correlation energy by scaling the Coulomb-like opposite-spin correlation component with an empirical factor, thereby avoiding the complexities associated with the exchange-like terms in the MP2 expression.\cite{jung2004scaled}  Avoiding the exchange-like terms makes it particularly appealing for proof-of-concept work, such as this.  In addition to exploiting the sparsity of the ERI tensor, our algorithm also leverages the inherent sparsity of the density matrices\cite{rudberg2008linearhf,rudberg2011lineardft,vandevondele2012millionscf} involved in the atomic-orbital Laplace transform implementation of MP2.\cite{haser1993theor}

In Section~\ref{sec:Background and Related Work}, we introduce essential background information related to atomic orbital Laplace transform MP2 and the $\mathcal{H}^2$-matrix representation.  In Section~\ref{sec:Methodology} we describe our hierarchical SOS-MP2 algorithm, and in Section~\ref{sec:Results} we demonstrate its accuracy and performance via numerical experiments.  Finally, in Section~\ref{sec:Conclusion} we draw conclusions and discuss future prospects.

\section{Background and Related Work}
\label{sec:Background and Related Work}
\subsection{Atomic Orbital Laplace MP2}
\label{subsec:AO-Laplace MP2}

In this section, we briefly review the atomic orbital Laplace MP2 method, first discussed in detail by Häser.\cite{haser1992laplace} The Coulomb-like term of the MP2 energy is given by
\begin{equation}
E_{2} = -2 \sum_{i,j,a,b} \frac{(ia|jb)^2}{\epsilon_{a} + \epsilon_{b} - \epsilon_{i} - \epsilon_{j}}.
\end{equation}
To facilitate the evaluation of the denominator, we perform a Laplace transformation and apply a quadrature procedure, introducing weights and abscissae, $\{w_{\alpha}, t_{\alpha}\}$,
\begin{equation}
E_{2} = -\sum_{\alpha}^{\tau} w_{\alpha} e_{2}^{\alpha}.
\end{equation}
In what follows, we focus on the evaluation of $e_{2}^{\alpha}$ at a single quadrature point, with the total SOS-MP2 energy obtained as a weighted sum over all points.

The Hartree--Fock energy-weighted density matrices and their complements are defined as
\begin{equation}
X_{\mu\nu}^{\alpha} = \sum_{i}^{\text{occ}} C_{\mu i} C_{\nu i} e^{\epsilon_{i} t_{\alpha}}, \quad
Y_{\mu\nu}^{\alpha} = \sum_{a}^{\text{vir}} C_{\mu a} C_{\nu a} e^{-\epsilon_{a} t_{\alpha}}.
\end{equation}
The electron repulsion integrals (ERIs) in this transformed index space are then expressed as
\begin{equation}
(\underline{\mu} \overline{\nu} | \underline{\lambda} \overline{\sigma}) = \sum_{\gamma\delta\kappa\epsilon} X_{\mu\gamma}^{\alpha} Y_{\nu\delta}^{\alpha} (\gamma\delta | \kappa\epsilon) X_{\kappa\lambda}^{\alpha} Y_{\epsilon\sigma}^{\alpha},
\end{equation}
where under- and overlined indices indicate transformation by $\mathbf{X}^\alpha$ and $\mathbf{Y}^\alpha$, respectively.
Accordingly, the MP2 energy contribution at quadrature point $\alpha$ is given by
\begin{equation}
e_{2}^{\alpha} = -2 \sum_{\mu\nu\lambda\sigma} (\underline{\mu^{\alpha}} \overline{\nu^{\alpha}} | \underline{\lambda^{\alpha}} \overline{\sigma^{\alpha}}) (\mu\nu | \lambda\sigma).
\end{equation}

\subsection{Hierarchical Approximation of ERI Tensor}
\label{subsec:H2ERI}

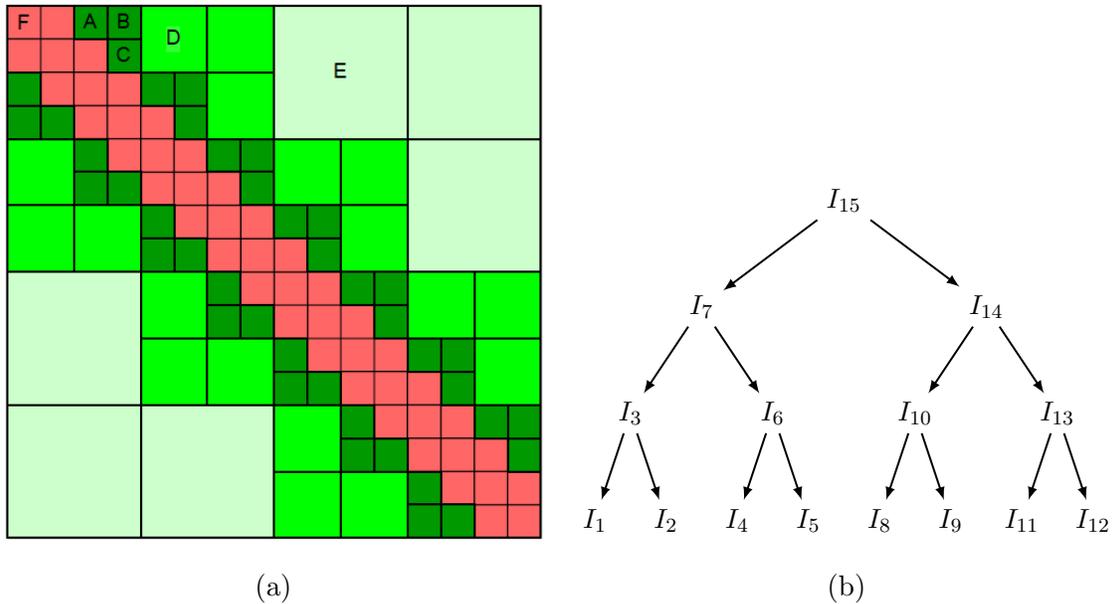
\begin{figure}[htbp]
  \centering
  \hfill
  \begin{subfigure}[t]{0.48\textwidth}
      \centering
      \includegraphics[width=\textwidth]{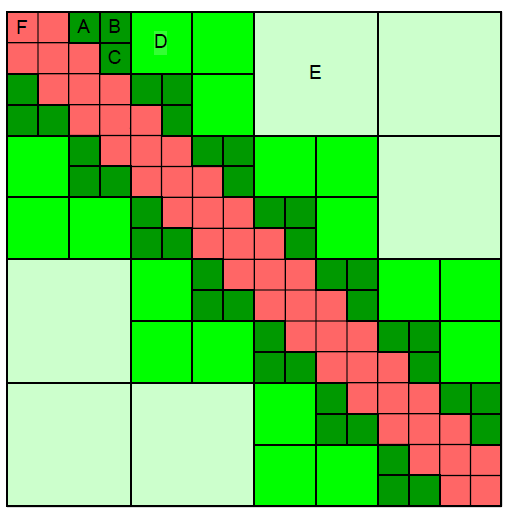}
      \subcaption{}
      \label{fig:h2}
  \end{subfigure}
  \hfill
  \begin{subfigure}[t]{0.48\textwidth}
     \raggedright
     \resizebox{\linewidth}{!}{%
       \begin{tikzpicture}[
         level 1/.style={sibling distance=40mm, level distance=15mm},
         level 2/.style={sibling distance=20mm, level distance=15mm},
         level 3/.style={sibling distance=10mm, level distance=15mm},
         edge from parent/.style={draw, -latex, thick},
         edge from parent path={(\tikzparentnode) -- (\tikzchildnode)},
         every node/.style={text height=1.5ex, text depth=.25ex}
        ]
         \node {$I_{15}$}
           child {node {$I_7$}
             child {node {$I_3$}
               child {node {$I_1$}}
               child {node {$I_2$}}
             }
            child {node {$I_6$}
               child {node {$I_4$}}
               child {node {$I_5$}}
            }
           }
           child {node {$I_{14}$}
             child {node {$I_{10}$}
               child {node {$I_8$}}
               child {node {$I_9$}}
             }
             child {node {$I_{13}$}
               child {node {$I_{11}$}}
               child {node {$I_{12}$}}
             }
           };
        \end{tikzpicture}%
      }
      \subcaption{}
      \label{fig:tree-structure}
  \end{subfigure}

  \caption{Illustration of (a) the $\mathcal{H}^2$ matrix representation of the ERI, and  (b) a hierarchical tree structure defining the blocking of rows and columns.}
  \label{fig:combined}
\end{figure}

In this section, we describe the hierarchical matrix formalism for ERI compression.  Consider the ERI tensor, $(\gamma\delta | \kappa\epsilon)$, to be an $N^2 \times N^2$ matrix, as illustrated in Figure~\ref{fig:h2}.  Each row or column corresponds to a particular basis function pair, e.g. $\gamma\delta$.  The ERI is partitioned into a hierarchy of sub-blocks of different sizes, with smaller blocks tending to be near the diagonal.  In a typical implementation of hierarchical matrices, the red blocks, such as F, would be stored in a dense form, and the off-diagonal green blocks would be compressed.  However, we achieve additional compression by storing the diagonal and near-diagonal red blocks in sparse form, approximating elements that fall below a user-chosen threshold as zero.  We refer to these blocks as \emph{short-range} blocks, because they represent electrostatic interactions at short range.  The green blocks of various sizes, which we will refer to as \emph{long-range} blocks, are compressed more aggressively than the short-range blocks, using a low-rank representation.  This strategy is designed to leverage the fact that the long-range blocks corresponding to distant atom pairs are inherently of low rank.

The blocking scheme is encoded by a tree structure that reflects physical space.  The rows and columns of the ERI matrix are each organized into such a tree, resulting in a decomposition of the matrix into the product of row and column trees. Figure~\ref{fig:tree-structure} illustrates a tree structure labeled in post-order traversal, where each node corresponds to a block containing several atomic basis function pairs in the system, and each child node denotes a sub-block contained within its parent. The nodes are partitioned such that the centroids of the basis function pairs associated with each node are near each other in physical space.  In the example shown in Figure~\ref{fig:tree-structure}, each parent node has two children for simplicity.  However, in three-dimensional systems, partitioning may occur in any subset of the three Cartesian directions; thus, a parent node may have up to eight children.

For a hierarchical matrix of rank $k$, the long-range blocks will be approximated to have rank at most $k$, and thus can be stored as $M = AB^{T}$, where $M$ is the block and $A$ and $B$ are rectangular matrices of rank $k$. By bounding $k$, the storage cost is reduced to storing only $A$ and $B$, resulting in an overall storage complexity of $\mathcal{O}(m \log m)$, where $m$ denotes the number of rows and columns.

The $\mathcal{H}^2$ matrix format provides further compression of hierarchical matrices by introducing nested row and column cluster trees.\cite{hackbusch2002data,huang2020h2pack} In the $\mathcal{H}^2$ format, two blocks in the same row or column also share the same row or column basis, respectively. Given that many blocks now share the same row and column bases, each block only needs to store a small \emph{intermediate matrix}, with the larger basis matrices stored only once per row/column.  We denote the intermediate matrix of block $K$ as $I_K$ and the row and column bases as $U_i$, which are the same in the case of a symmetric matrix. Then, for example, the dark green matrices can be expressed as $A = U_1 I_A U_4^{T}$, $B = U_1 I_B U_5^{T}$, and $C = U_2 I_C U_5^{T}$. Note that A and B share the same row basis, $U_1$, because they are in the same row, and B and C share the same column basis, $U_5$, because they are in the same column.

As one moves farther from the diagonal, the blocks become larger, but $k$ remains unchanged, so the compression ratio increases.  Moreover, if the row (column) of one block is a subset of the row (column) of another, their row (column) bases can be related via a \emph{transfer matrix}, eliminating the need to store a new basis for these larger blocks.  For example, the matrix $D$ is written as $D = U_3 I_D U_{10}^{T}$. As shown in Figure~\ref{fig:tree-structure}, node $3$ is the parent of nodes $1$ and $2$. The nested row basis can be written as
\[
U_3 = \begin{bmatrix} U_1 & 0 \\ 0 & U_2 \end{bmatrix} \begin{bmatrix} R_1 \\ R_2 \end{bmatrix},
\]
where $R_i$ denotes the transfer matrix of node $i$.
The row and column bases of larger blocks, such as $E$, can be computed recursively, maintaining a constant storage cost per block. Asymptotically, the $\mathcal{H}^2$ format achieves linear storage complexity.  This is because only the row/column bases at the leaf nodes need be stored explicitly, while the intermediate matrices and the transfer matrices collectively require $\mathcal{O}(m)$ storage. For $\mathcal{H}^2$ compression of the ERI tensor, $m$ denotes the number of screened basis function pairs, satisfying $m \leq N^{2}$ for a system with $N$ basis functions. In the complexity analysis presented below, we use $m = \mathcal{O}(N^{2})$.  Hereafter, we refer to column/row bases that are preserved between blocks and inherited across levels of the hierarchy as \emph{nested}.

\section{Methodology}
\label{sec:Methodology}
Here we describe the key steps in our hierarchical SOS-MP2 algorithm.  The overall procedure is presented in Algorithm~\ref{alg:HSM algorithm}.

\renewcommand{\algorithmicrequire}{\textbf{Input:}}
\renewcommand{\algorithmicensure}{\textbf{Output:}}
\noindent\rule{\linewidth}{0.8pt} 
\vspace{-30pt} 
\captionof{algorithm}{Hierarchical SOS-MP2 algorithm}
\label{alg:HSM algorithm}
\vspace{-15pt} 
\noindent\rule{\linewidth}{0.4pt} 

\begin{algorithmic}[1]
\Require The atomic orbital ERI tensor $W$ in $\mathcal{H}^2$ format. The coefficient matrix $C$ and occupied and virtual molecular orbitals $e_i$. The quadrature weights and abscissa $\{w_{\alpha}, t_{\alpha}\}$
\Ensure MP2 Coulomb-like term energy $E_{2,J}$
\For{$\alpha$ in quadrature points}
\State Calculate $X_{\mu\nu}^{\alpha} = \sum_{i}^{\text{occ}} C_{\mu i}C_{\nu i} e^{\epsilon_{i} t_{\alpha}}$ and  $Y_{\mu\nu}^{\alpha} = \sum_{a}^{\text{vir}} C_{\mu a} C_{\nu a} e^{-\epsilon_{a} t_{\alpha}}$
\State Do $X$ and $Y$ index transformations on $W_s$ to get transformed short-range part $T_s$  
\For{Block $B_i$ in completely low-rank format}
\For{Column $\kappa\epsilon$ in the block}
\For{$\lambda$ where $X_{\kappa\lambda}$ is the truncated significant elements}
\State Compare the block containing column $\lambda\epsilon$ with $B_i$
\If {They are in the same level}
\State {Directly product the column basis set using \eqref{eq:case1}}
\ElsIf{the transformed block is of lower level}
\State{Do ancestor index transformation using \eqref{eq:case2}}
\Else
\State{Do descendant index transformation using \eqref{eq:case3}}
\EndIf
\EndFor
\EndFor
\EndFor
\State Get the half index transformed low-rank part ERI tensor $(\mu \nu|\underline{\lambda} \epsilon)$
\State Similarly do $Y$ index transformation to get transformed long-range part $T_l=(\mu \nu|\underline{\lambda} \overline{\sigma})$
\State Loop over all the elements of $T_s$ to compute $\text{Tr}(T_s^2+2T_s T_l)$.
\State Loop over all the blocks of $T_l$ to compute $\text{Tr}(T_l^2)$.
\State Summarize this quadrature point: $e_{2}^{\alpha} = -2 \, \text{Tr}\left[(T_s^2) + (T_l^2) + 2 T_s \cdot T_l\right]$
\EndFor
\State \Return $E_2^{\text{SOS-MP2}}=\sum_{\alpha} \omega_{\alpha}e^{\alpha}_2$
\end{algorithmic}
\vspace{-5pt} 
\noindent\rule{\linewidth}{0.8pt} 

\subsection{Partitioning of the ERI Tensor}
\label{subsec:Decomposition of The ERI Tensor}
The SCF procedure involves the multiplication of the ERI tensor by the density matrix, which can be understood as a matrix-vector product in the space of basis function pairs.  Such an operation is well-suited to the $\mathcal{H}^2$-matrix representation. By contrast, the SOS-MP2 method involves operations that resemble a matrix-matrix product, which is less ideal for the direct application of the $\mathcal{H}^2$ format. Nonetheless, for finite systems, the electron density decays exponentially with distance, and insulator-type systems exhibit the fastest decay.\cite{kohn1995density,kohn1996density} As a result, the density and density complementary matrices are expected to be sparse, which allows us to avoid a full, complex matrix-matrix multiplication. Furthermore, substantial cancellations observed in MP2 calculations suggest that accurate results can be achieved by focusing on the dominant elements in one matrix, thereby further simplifying the computation.

In what follows, we assume that the SCF phase of the calculation is complete, and therefore the orbital coefficient matrices, $C$, are already available. With $C$, the energy-weighted density and density complementary matrices, $X$ and $Y$, can be computed as
\begin{equation}
X_{\mu\nu}^{\alpha} = \sum_{i}^{\text{occ}} C_{\mu i} C_{\nu i} e^{\epsilon_{i} t_{\alpha}}
\end{equation}
and
\begin{equation}
\quad Y_{\mu\nu}^{\alpha} = \sum_{a}^{\text{vir}} C_{\mu a} C_{\nu a} e^{-\epsilon_{a} t_{\alpha}}.
\end{equation}
Both calculations have a time complexity of $\mathcal{O}(N^3)$. However, these computations are performed only once, and the associated prefactor is sufficiently small that the overall time complexity is negligible for the systems considered in this work.  $X$ and $Y$ are both sparse in larger systems.  We therefore store them in compressed sparse row (CSR) format.  Elements below a user-chosen threshold, $\eta$, are approximated as zero.

As described in Section~\ref{subsec:AO-Laplace MP2}, the Coulomb-like term of the MP2 energy can be expressed as a weighted sum over quadrature points, $E_{2} = -\sum_{\alpha}^{\tau} w_{\alpha} e_{2}^{\alpha}$. For clarity, we focus on the calculation of $e_{2}^{\alpha}$ at a single quadrature point and omit the superscript $\alpha$ in the derivation that follows. In Section~\ref{sec:Results}, we employ the quadrature rules developed by Braess and Hackbusch.\cite{braess2005approximation,takatsuka2008minimax} Typically, seven or eight quadrature points are used.

The Coulomb-like term of the MP2 energy can be written explicitly as
\begin{equation}
  e_{2} = -2 \sum_{\mu,\nu,\lambda,\sigma,\gamma,\delta,\kappa,\epsilon} (\mu \nu | \lambda \sigma) X_{\mu\gamma} Y_{\nu \delta} (\gamma \delta | \kappa \epsilon) X_{\kappa \lambda} Y_{\epsilon \sigma}.
\end{equation}
Introducing $W$ as the ERI tensor, we can express the equation more compactly as
\begin{equation}
  e_{2} = -2 \, \text{Tr}\left[(W (X \otimes Y))^2\right],
\end{equation}
where $\text{Tr}$ denotes the trace and $\otimes$ denotes the Kronecker product. Defining
\begin{equation}
T = W (X \otimes Y),
\end{equation}
the energy expression simplifies further to
\begin{equation}
    e_{2} = -2 \, \text{Tr}(T^2).
\end{equation}

We refer to the operation of multiplying $W$ by $X \otimes Y$ as \emph{index transformation}. Here, $W$ denotes the original ERI tensor, while $T$ denotes the index-transformed ERI tensor. As discussed in Section~\ref{subsec:H2ERI}, the ERI tensor, represented as an $\mathcal{H}^2$-matrix, can be decomposed into a short-range component, $W_s$, which includes diagonal and near-diagonal blocks, and a long-range component, $W_l$, including the remainder of the matrix. Owing to the linearity of matrix-matrix multiplication,
\begin{equation}
    T=T_s+T_l
\end{equation}
where
\begin{equation}
\label{eq:shortrangeindextransform}
    T_s = W_s (X \otimes Y)
\end{equation}
and
\begin{equation}
\label{eq:longrangeindextransform}
    T_l = W_l (X \otimes Y).
\end{equation}

The Coulomb-like term of the MP2 energy then takes the form
\begin{equation}
\label{eq:e2}
e_{2} = -2 \, \text{Tr}\left[T_s^2 + T_l^2 + 2 T_s T_l\right].
\end{equation}
It is important to note that $T_s$ and $T_l$ do not denote the short- and long-range parts of the index-transformed ERI tensor.  Instead, they are the index-transformed versions of the short- and long-range parts of the original ERI tensor.

\subsection{Index Transformation of the Short-Range Component}
\label{subsec:Index Transformation of the dense part}
Here we describe the transformation of the short-range component of the ERI tensor, Eq.~\eqref{eq:shortrangeindextransform}.
Both the original $W_s$ and the ultimate index-transformed $T_s$ are stored in CSR matrix format in order to leverage their sparsity, with elements whose absolute values fall below a second sparsification threshold, $\zeta$, are approximated as zero.
In practice, the transformation is carried out in two steps, applying $X$ and $Y$ to the ERI tensor separately. First, the $X$ index transformation is computed as
\begin{equation}
(\mu \nu | \underline{\lambda} \epsilon) = (\mu \nu | \kappa \epsilon) X_{\kappa \lambda}.
\end{equation}
Following this transformation, the resulting intermediate ERI tensor is sparsified by applying the threshold $\zeta$.
 The $Y$ index transformation is then performed analogously
 \begin{equation}
 (\mu \nu | \underline{\lambda} \overline{\sigma}) = (\mu \nu | \underline{\lambda} \epsilon) Y_{\epsilon \sigma}.
 \end{equation}

A naive implementation of this step would scale as $\mathcal{O}(N^5)$. However, the sparsity of $X$ and $Y$ results in reduced scaling.  In large systems, the number of significant elements of $X$ and $Y$ scales as $\mathcal{O}(N)$.  Applying only these significant elements, the time and space complexities of these short-range index transformation steps scale as $\mathcal{O}(N^2)$. Another factor contributing to the efficiency of the storage is the substantial cancellation observed in the index transformation of the short-range component, although a rigorous mathematical proof of this cancellation has not yet been established.

\subsection{Index Transformation of the Long-Range Component}
\label{subsec:Index Transformation of the low rank part}

Now we discuss the index transformation of the long-range component, Eq.~\eqref{eq:longrangeindextransform}.
The long-range part of the ERI, $W_l$, comprises a set of low-rank matrix blocks in an $\mathcal{H}^2$ representation. Since this index transformation applies a right multiplication to $W_l$, the nested property of the row basis sets is preserved, while the nested property in the column basis sets is lost.  Nevertheless, the row basis sets of blocks in the same row of blocks remain connected via transfer matrices.

We begin by converting the $\mathcal{H}^2$ matrix into a more general hierarchical matrix, no longer enforcing the nested property of the column basis.  This is done by retaining the row basis sets and multiplying the intermediate matrices by the column basis sets. This step ensures that the nested property of the row basis sets is maintained across different blocks. The time and space complexities of this step both scale as $\mathcal{O}(N^2)$.

Next, we perform the $X$ index transformation on the hierarchical matrix by evaluating
\[
(\mu \nu | \underline{\lambda} \epsilon) = (\mu \nu | \kappa \epsilon) X_{\kappa \lambda},
\]
and store $(\mu \nu | \underline{\lambda} \epsilon)$ in a \emph{completely low-rank hierarchical matrix} format. Such a matrix is partitioned in the same manner as in the original hierarchical matrix.  However, all blocks, including the diagonal and neighboring blocks, are stored in low-rank format. Because right multiplication does not affect the row basis, each block’s row basis set in $(\mu \nu | \underline{\lambda} \epsilon)$ matches that in $(\mu \nu | \kappa \epsilon)$. We then iterate over all blocks in $(\mu \nu | \underline{\lambda} \epsilon)$, including both diagonal and neighboring blocks, to compute their column basis sets. Each column basis represents a basis function pair $\underline{\lambda} \epsilon$. When determining the influence of the row elements, $X_{\kappa \lambda}$, on the column basis sets, we consider three cases based on the relationship between the row basis sets:

\begin{description}[leftmargin=0pt, labelwidth=\widthof{\textbf{Descendant Index Transformation}}]

  \item[\textbf{Same-level Index Transformation}] \par
  The row basis set of the block containing $\kappa \epsilon$ is identical to that of the block containing $\underline{\lambda} \epsilon$.

  \item[\textbf{Ancestor Index Transformation}] \par
  The row basis set of the block containing $\kappa \epsilon$ is a subset of the row basis set of the block containing $\underline{\lambda} \epsilon$.

  \item[\textbf{Descendant Index Transformation}] \par
  The row basis set of the block containing $\kappa \epsilon$ is a superset of the row basis set of the block containing $\underline{\lambda} \epsilon$.

\end{description}

Let $y$ denote the column basis vector $\underline{\lambda} \epsilon$ to be computed, and let $x$ denote the column basis set in $\kappa \epsilon$. When the block containing $\kappa \epsilon$ lies in the short-range part, it is treated as an empty block, as this part has already been computed in Section~\ref{subsec:Index Transformation of the dense part}.

In the same-level transformation, we directly multiply the value $X_{\kappa \lambda}$ with the column basis vector $x$ to obtain the contribution to $y$, since the row basis sets of both blocks are identical. This can be written as
\begin{equation}
  y = X_{\kappa \lambda} x.
  \label{eq:case1}
\end{equation}
The time complexity of each individual transformation is $\mathcal{O}(1)$. This step is implemented in line 9 of Algorithm~\ref{alg:HSM algorithm}.

In the ancestor transformation, we trace the sequence of ancestors $n_i$ in the row tree from the block containing $\kappa \epsilon$ to the block containing $\underline{\lambda} \epsilon$. The row basis set of the block containing $\kappa \epsilon$ is effectively the recursive product of the transfer matrices along this ancestor sequence. We express this as
\begin{equation}
  y = X_{\kappa \lambda} \prod_{i} R_{n_i} x.
  \label{eq:case2}
\end{equation}
The time complexity of this process is $\mathcal{O}(\log N)$ per transformation, as there are at most $\mathcal{O}(\log N)$ levels in the tree.  For large systems, this level difference is typically small, since $X$ is expected to be sparse with its most significant values near the diagonal. This step is implemented in line 11 of Algorithm~\ref{alg:HSM algorithm}.

In the descendant transformation, the procedure is more involved. First, we identify the blocks containing the required column $\kappa \epsilon$, denoted by a set of blocks, ${B_i}$, with corresponding row basis sets ${U_i}$ and columns ${x_i}$. For each $U_i$, we then trace a sequence of ancestors, $n_{ij}$, in the row tree from the block containing $\underline{\lambda} \epsilon$ down to the block containing $U_i$. This sequence is the reverse of that in the ancestor transformation, moving from descendant to ancestor rather than vice versa. We write
\begin{equation}
  y = \sum_{i} X_{\kappa \lambda} \prod_{j} R^{-1}_{n_{ij}} x_i,
  \label{eq:case3}
\end{equation}
where $R^{-1}_{n_{ij}}$ denotes the pseudo-inverse of the transfer matrix, which is precomputed with a time complexity of $\mathcal{O}(N^2 \log N)$. The time complexity of each transformation itself is $\mathcal{O}(\log^2 N)$.  However, this worst-case behavior is rarely encountered in practice, typically arising only when computing interactions between distant basis function pairs. As noted previously, significant elements of $X$ are generally concentrated near the diagonal.

Approximately $\mathcal{O}(N^2 \log N)$ column basis sets must be computed, representing the storage requirement of this step. If $X$ were dense, the total time complexity would be $\mathcal{O}(N^2 \log^3 N)$---$\mathcal{O}(\log^2 N)$ for each of the $\mathcal{O}(N^2 \log N)$ column basis sets. However, in large systems, $X$ is expected to be sparse.  Because more distant basis function pairs are more likely to correspond to zero elements in $X$, same-level transformations are much more common in practice than ancestor or descendant transformations.  Thus, the same-level transformation dominates the time, leading to an effective time complexity of $\mathcal{O}(N^2 \log N)$. This step is implemented in line 13 of Algorithm~\ref{alg:HSM algorithm}.

The $Y$ index transformation step proceeds analogously to the $X$ transformation step. The key difference is that diagonal and neighboring blocks are no longer treated as empty but instead are represented as low-rank matrices sharing the same row basis sets as their neighbors. The time and space complexities for the $Y$ transformation are also $\mathcal{O}(N^2 \log N)$ for large systems. In practice, the error introduced by the long-range index transformation is found to be smaller than that of the short-range transformation, which allows a higher threshold to be used when sparsifying $X$ and $Y$ compared to the threshold used for the short-range part of the index transformation.

\subsection{Computation of MP2 Energy}
\label{subsec:Computation of MP2 Energy}
With the CSR matrix, $T_{\text{s}}$, and the completely low-rank hierarchical matrix, $T_{\text{l}}$, representing the short- and long-range components of the index-transformed ERI tensor, the Coulomb-like term of the MP2 energy is given by \eqref{eq:e2}.
The terms $\text{Tr}(T_{\text{s}}^2)$ and $\text{Tr}(T_{\text{s}} T_{\text{l}})$ are evaluated directly by iterating over all elements of $T_{\text{s}}$ and identifying the corresponding elements in the other matrix to compute their contributions to the total trace. Both steps have a time complexity of $\mathcal{O}(N^2)$.

For the term $\text{Tr}(T_{\text{l}}^2)$, due to the symmetric block structure of the hierarchical matrix, it is sufficient to select each block and compute the trace of the product of the block with its mirror image across the diagonal. Since both blocks are stored in low-rank format, this step has a time complexity of $\mathcal{O}(N^2 \log N)$.  Consequently, the overall time complexity for computing the SOS-MP2 energy is $\mathcal{O}(N^2 \log N)$.

\subsection{Error Analysis} 
\label{subsec:Error Analysis}

Now we turn our attention to analyzing the sources of numerical error in our algorithm.  The errors in the index transformation procedure primarily arise from neglecting small elements of $X$ and $Y$ to enable sparse storage. Let $X_{\text{r}}$ and $Y_{\text{r}}$ denote the truncated matrices. The resulting error in $T_{\text{s}}$ comprises the terms
\begin{equation}
T_{\text{sX}} = W_{\text{d}} (X_{\text{r}} \otimes Y),
\end{equation}
\begin{equation}
T_{\text{sY}} = W_{\text{d}} (X \otimes Y_{\text{r}}),
\end{equation}
and
\begin{equation}
T_{\text{sXY}} = W_{\text{d}} (X_{\text{r}} \otimes Y_{\text{r}}),
\end{equation}
where subscript $\text{r}$ indicates the residual after sparsification.
Analogous considerations apply to $T_{\text{l}}$. This error is similar to rounding error in numerical computations and can be controlled by selecting an appropriate threshold. The error introduced by the long-range index transformation is even smaller than that of the short-range index transformation, which permits the use of a higher threshold for $X$ and $Y$ in the short-range transformation step. If we denote $T_{\text{sX}}$ and $T_{\text{sY}}$ as $\mathcal{O}(\epsilon)$, the contribution from $T_{\text{sXY}}$ is of order $\mathcal{O}(\epsilon^2)$, which is negligible. In principle, the error could be further reduced from $\mathcal{O}(\epsilon)$ to $\mathcal{O}(\epsilon^2)$ by evaluating $\text{Tr}\left[T_{\text{s}}(T_{\text{sX}} + T_{\text{sY}})\right]$, which would add only an $\mathcal{O}(N^2)$ time complexity. However, in our numerical experiments, the observed error is already sufficiently small and comparable in magnitude to the error in $\text{Tr}(T_{\text{s}}^2)$, which cannot be evaluated as efficiently. As a result, this correction step was not included in the algorithm.

Another source of error is the low-rank approximation of the ERI tensor and the low-rank index transformation. Apart from the error introduced by thresholding, the low-rank approximation only contributes to the error in the descendant transformation step, where the pseudo-inverse of the transfer matrices is computed. This error contribution is found to be negligible in practice.

An additional source of error is the numerical error introduced by the Laplace transformation. However, the Laplace transformation is a well-established numerical technique, and the error it introduces is generally small enough to be neglected.

\subsection{Potential for Parallel Implementation}
\label{subsec:parallel properties}
In quantum chemistry, along with physical approximations and efficient numerical methods, parallelization is a key strategy to extend the size and complexity of systems that may be studied.\cite{calvin2021chemrev,bernholdt1996largescale,fletcher1999parallel,hattig2006distributedmemory,ufimtsev2008gpu1,Fales2015,chow2015parallel,werner2015scalable,peng2016tiledarray,kowalski2021nwchemex,hu2024das}  That hierarchical matrices naturally map to massively parallel computer architectures is a significant advantage that will be exploited in future work.  In Laplace transform MP2, the computation of the MP2 energy at each quadrature point is independent, making the algorithm highly parallelizable. Furthermore, each row computation in $T_{\text{s}}$ and each block computation in $T_{\text{l}}$ are also independent, enabling parallelization of the short- and long-range parts of the index transformation, respectively. This same property applies to the MP2 energy computation step. In other words, the Hierarchical SOS-MP2 algorithm could theoretically achieve constant time complexity with an infinite number of processors. However, since parallelization requires considerable memory, we implemented parallelization only for the quadrature points in this proof-of-concept paper. Future work will involve parallelizing other parts of the algorithm on massively parallel computers.

\section{Results}
\label{sec:Results}
All computations were carried out on the Seawulf cluster 
at Stony Brook University. We selected two model systems to study: an alkane chain system, which is representative of one-dimensional systems, and a water cluster system, which is representative of fully three-dimensional systems. All SOS-MP2 calculations were performed using the cc-pVDZ basis set. The threshold used for sparsification of both the short-range part of the ERI tensor and the index-transformed ERI tensor was set to $\zeta=1 \times 10^{-6}$.

Different thresholds for sparsification of $X$ and $Y$ are used in the short- and long-range transformations, labeled $\eta_s$ and $\eta_l$, respectively.  Except as noted below, we set the short-range truncation threshold to $\eta_s = 10^{-5}$ for the alkane chain systems and $\eta_s = 3 \times 10^{-4}$ for the water cluster.  For the long-range transformation, thresholds are $\eta_l = 10^{-4}$ for the alkane system and $\eta_l = 3 \times 10^{-4}$ for the water cluster.

\subsection{The Sparsity of Density and Density Complementary Matrices}
\label{subsec:The sparsity of density and density complementary matrices}

\begin{figure}[htbp]
    \centering
    \begin{minipage}[t]{0.48\textwidth}
        \centering
        \includegraphics[width=0.94\columnwidth]{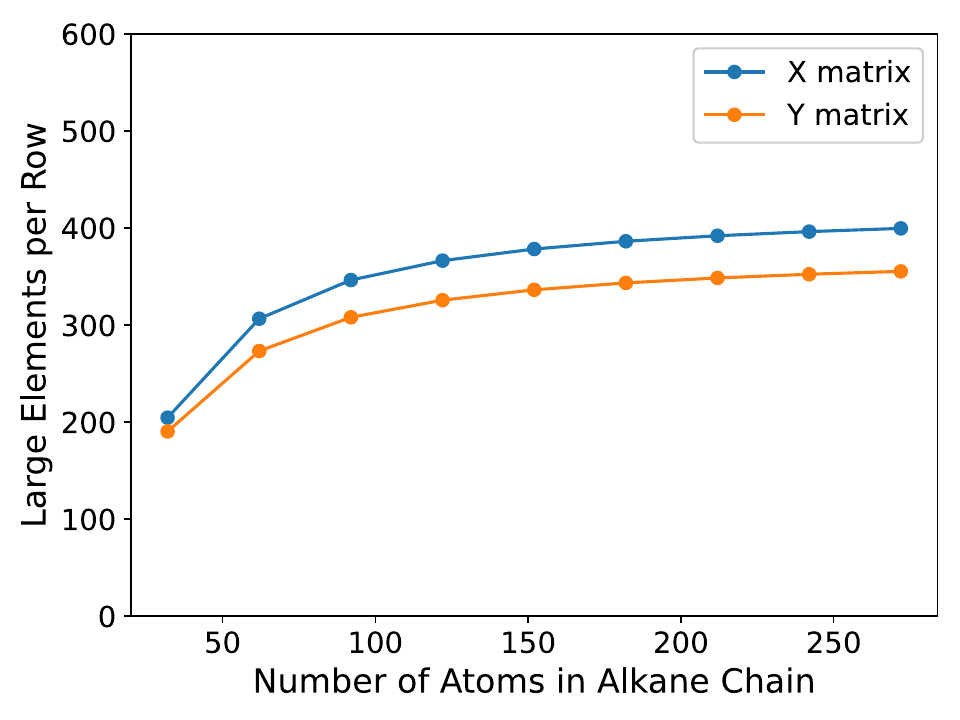}
        \caption{The number of significant values per row or column in the energy-weighted density matrices in the series of linear alkane chains.}
        \label{fig:sparsityalkane}
    \end{minipage}
    \hfill
    \begin{minipage}[t]{0.48\textwidth}
        \centering
        \includegraphics[width=0.94\columnwidth]{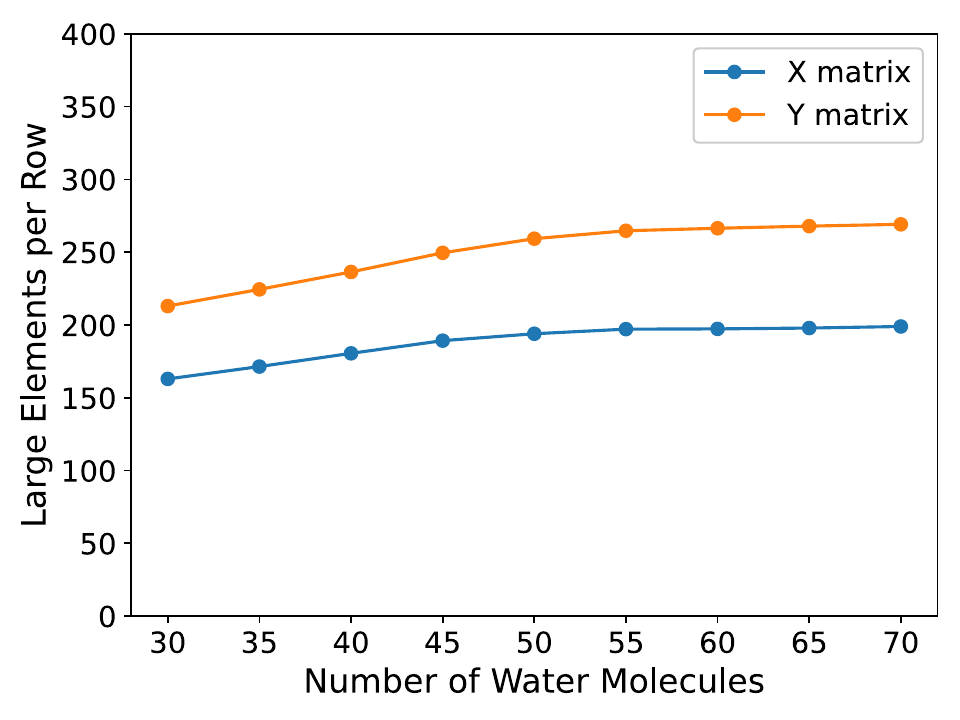}
        \caption{The number of significant values per row or column in the energy-weighted density matrices in the series of water clusters.}
        \label{fig:sparsitywater}
    \end{minipage}
\end{figure}

The performance of our algorithm is contingent on the sparsity of the energy-weighted density matrices $X$ and the density complementary matrices $Y$, thus we quantify that sparsity here.  We report the number of significant elements per row in Figure~\ref{fig:sparsityalkane} for the alkane chain and Figure~\ref{fig:sparsitywater} for the water cluster. Values were computed across all quadrature points, and the maximum value was retained. For the alkane system, we observe rapid growth in the number of significant values from $\mathrm{C}_{10}\mathrm{H}_{32}$ to $\mathrm{C}_{30}\mathrm{H}_{92}$. Beyond this range, growth slows and eventually saturates, consistent with the expected scaling of $\mathcal{O}(N)$ total significant density matrix element for large systems. The water cluster exhibits similar behavior---as the system size increases, the number of significant values per row reaches a plateau. These results confirm the asymptotic sparsity of both the density and complementary matrices in both 1-D and 3-D systems.

\begin{figure}[htbp]
  \centering
  \begin{minipage}[t]{0.48\textwidth}
      \centering
      \includegraphics[width=0.94\columnwidth]{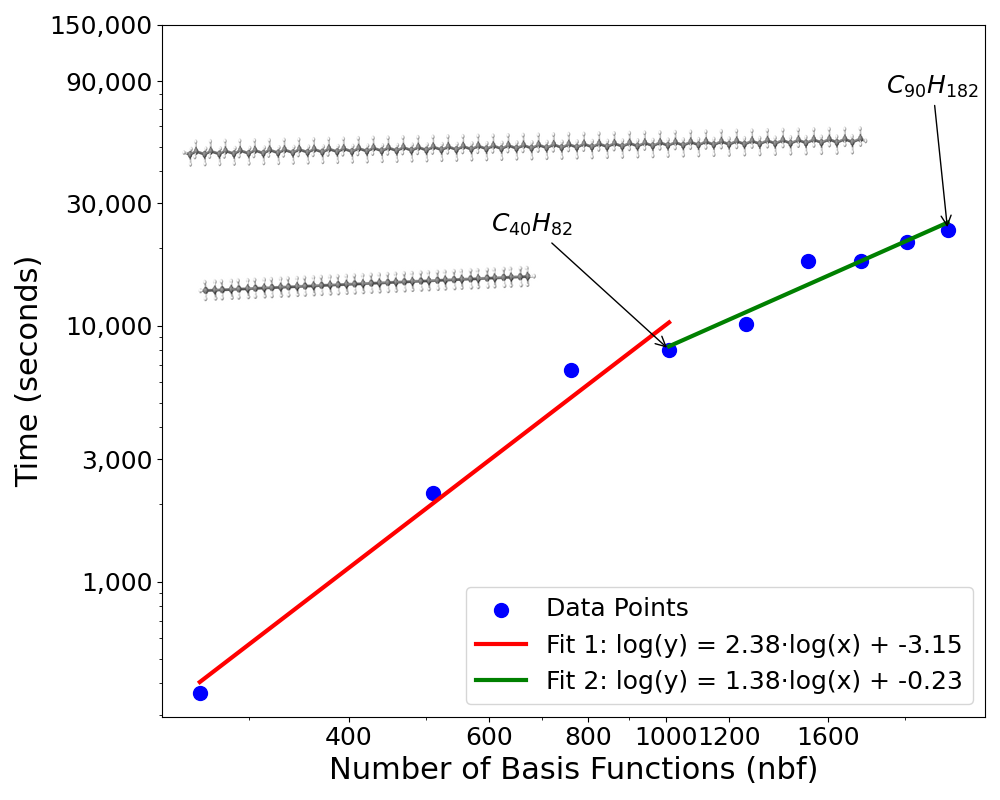}
      \caption{The time-to-solution versus number of basis functions for the series of linear alkane chains.}
      \label{fig:alktime}
  \end{minipage}
  \hfill
  \begin{minipage}[t]{0.48\textwidth}
      \centering
      \includegraphics[width=0.94\columnwidth]{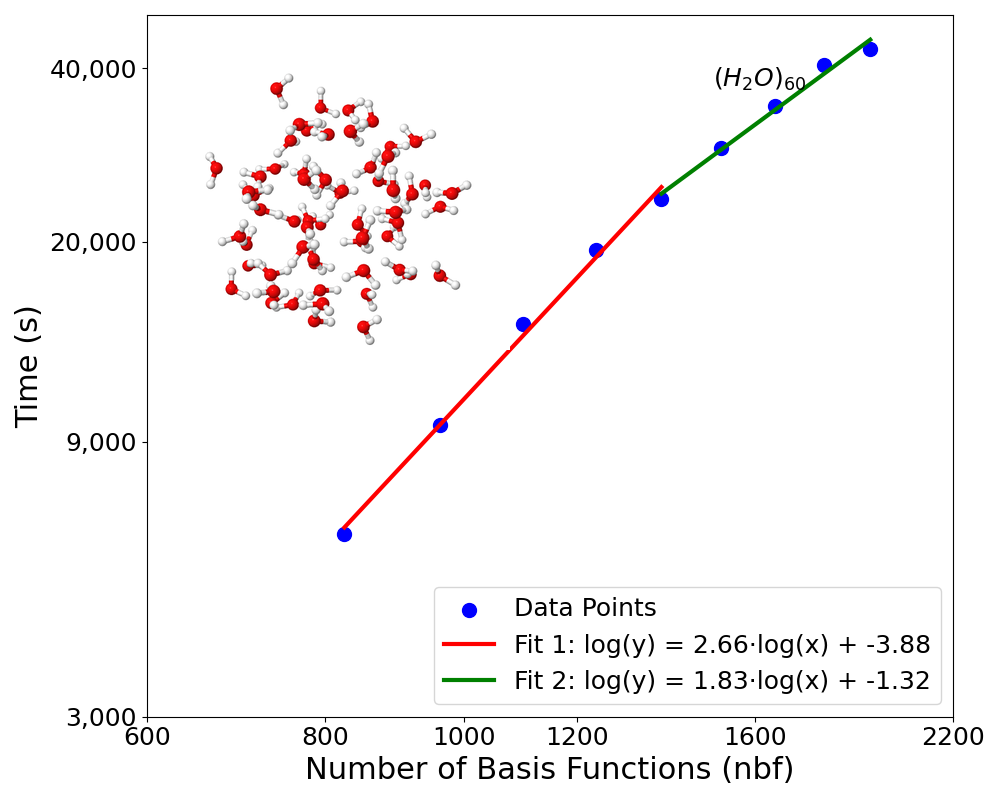}
      \caption{The time-to-solution versus number of basis functions for the series of water clusters.}
      \label{fig:watertime}
  \end{minipage}
\end{figure}

\subsection{Time Complexity}
\label{subsec:Time complexity analysis}
The measured times-to-solution for a series of linear alkane chains is shown in Figure~\ref{fig:alktime}. All timing results are averaged over three runs. Molecular structures were rendered using Jmol.\cite{Jmol}

The first four data points correspond to systems $\mathrm{C}_{10}\mathrm{H}_{22}$ through $\mathrm{C}_{40}\mathrm{H}_{82}$, in which the $X$ and $Y$ matrices have not yet reached asymptotic sparsity. Beyond this point, the number of nonzero elements per row of $X$ and $Y$ stabilizes. To illustrate the role of sparsity, we include two trend lines. For sufficiently large systems, we expect time complexity to scale as $\mathcal{O}(N^2 \log N)$, though the observed performance is closer to $\mathcal{O}(N^{1.38})$.  The superior performance is likely due to Schwarz screening. For smaller systems, complexity grows more quickly, which is consistent with expectations. The observed $\mathcal{O}(N^{2.38})$ scaling is still better than the theoretical $\mathcal{O}(N^3 \log^3 N)$ bound.  A discontinuity is observed between $\mathrm{C}_{50}\mathrm{H}_{152}$ and $\mathrm{C}_{60}\mathrm{H}_{182}$, corresponding to an increase in the number of levels in the hierarchical tree. Despite this, the increase in time-to-solution remains nearly linear.

Times-to-solution for the water cluster system are shown in Figure~\ref{fig:watertime}. For clusters with 30--50 molecules (first five points), time complexity scales approximately as $\mathcal{O}(N^{2.66})$. For clusters with 50--70 molecules (last five points), scaling improves to $\mathcal{O}(N^{1.83})$. The absolute time cost is higher for the water clusters than for the alkanes, as expected for a 3-D system, in which a larger fraction of the interactions are short range.

It is worth noting that, for the largest water clusters, we modified the hierarchical block-splitting algorithm relative to what was used for smaller clusters and alkane chains. In 3-D systems, the number of interacting block pairs grows rapidly with system size. To mitigate this, we cap the maximum depth of the block tree to control the number of interactions. This change improves performance and preserves the desired asymptotic behavior.

\begin{figure}[htbp]
  \centering
  \begin{minipage}[t]{0.49\textwidth}
      \centering
      \includegraphics[width=0.94\columnwidth]{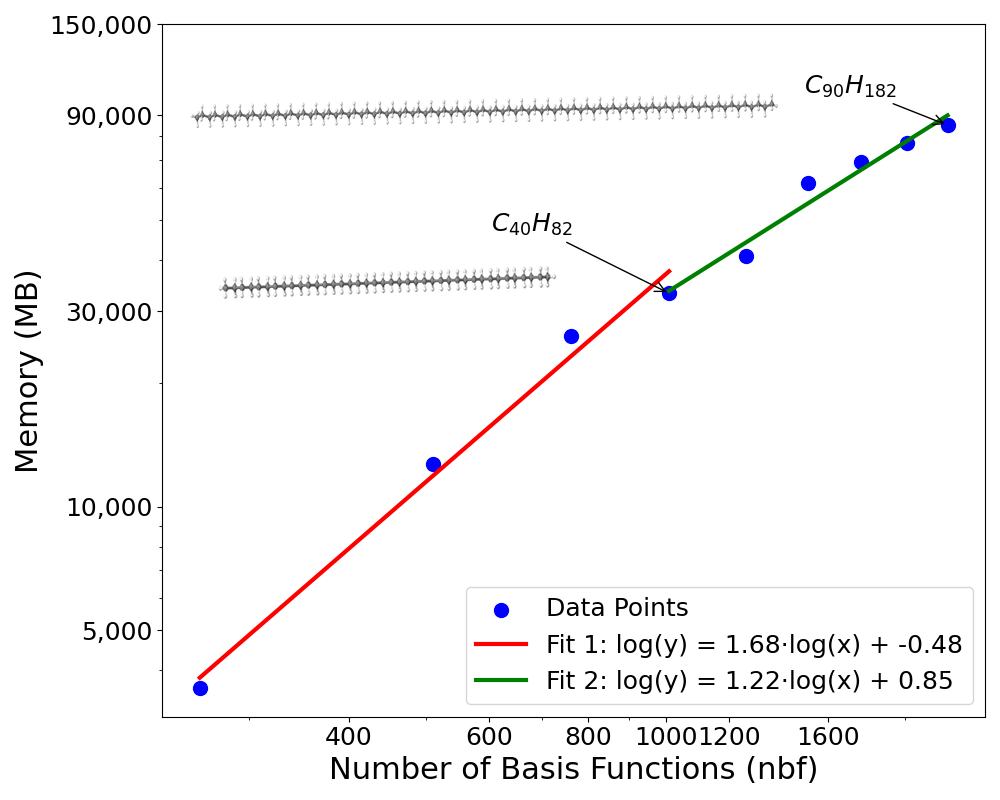}
      \caption{The memory versus number of basis functions for the series of linear alkane chains.}
      \label{fig:alkmem}
  \end{minipage}
  \hfill
  \begin{minipage}[t]{0.49\textwidth}
      \centering
      \includegraphics[width=0.94\columnwidth]{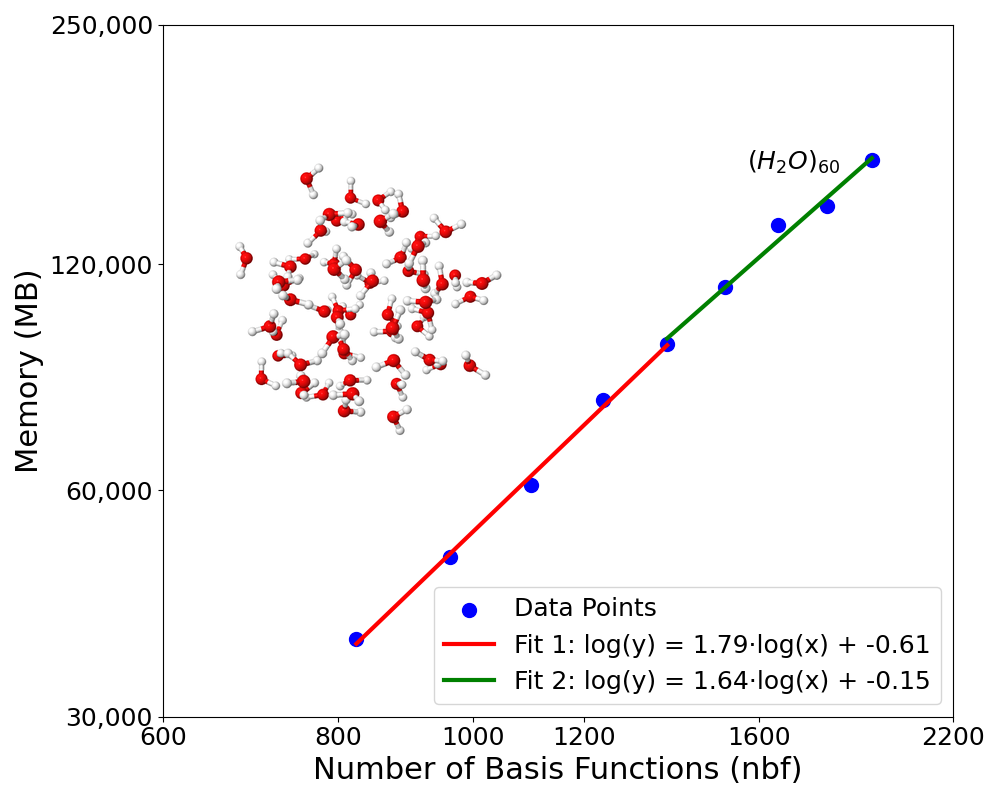}
      \caption{The memory versus number of basis functions for the series of water clusters.}
      \label{fig:watermem}
  \end{minipage}
\end{figure}

\subsection{Space Complexity}
\label{subsec:Space complexity analysis}

The total memory storage needed for the alkane chain systems is shown in Figure~\ref{fig:alkmem}. Their behavior closely parallels the time complexity results, though the rate of growth is even slower. For systems smaller than $\mathrm{C}_{40}\mathrm{H}_{82}$, where $X$ and $Y$ have not yet reached asymptotic sparsity, storage scales as $\mathcal{O}(N^{1.68})$. For larger systems, the theoretical scaling is $\mathcal{O}(N^2 \log N)$, but observed performance is $\mathcal{O}(N^{1.22})$, approaching linear. As in the time complexity case, we observe a step increase between $\mathrm{C}_{50}\mathrm{H}_{102}$ and $\mathrm{C}_{60}\mathrm{H}_{122}$, again due to an increase in the number of levels in the hierarchical tree.

Figure~\ref{fig:watermem} presents the memory storage for the water clusters. For clusters with 30--50 molecules, the observed scaling is approximately $\mathcal{O}(N^{1.79})$, while storage scales as $\mathcal{O}(N^{1.64})$ for clusters with 50--70 molecules. In contrast to the alkane chain system, this scaling does not flatten substantially with increasing system size, but remains within acceptable bounds and significantly better than the theoretical bound.

We attribute the slowdown in the alkane chain's asymptotic growth to Schwarz screening. In 1-D systems, the number of atoms within a given radius of a particular atom is small, and the majority of atom pairs interact weakly. Schwarz screening is therefore highly effective in eliminating small terms, yielding reduced memory costs. In 3-D systems, more pairs of atoms are close enough to interact strongly, leaving fewer terms that can be ignored. As a result, memory cost does not flatten as dramatically.

\begin{figure}[htbp]
  \centering
  \begin{minipage}[t]{0.48\textwidth}
     \centering
     \includegraphics[width=0.94\columnwidth]{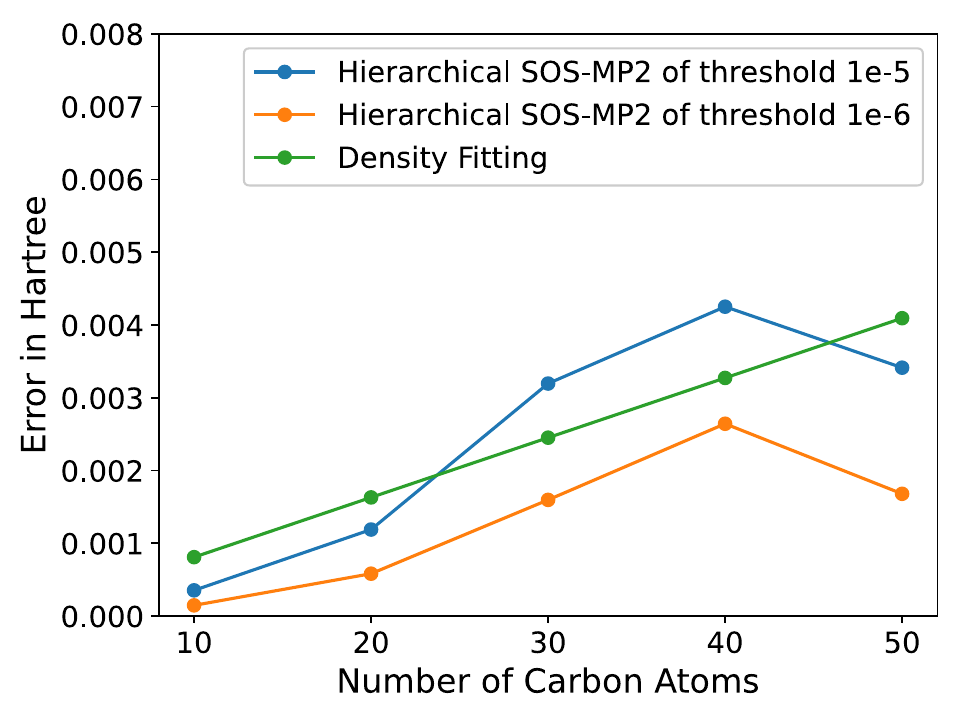}
     \caption{The error for alkane chain, computed with default thresholds ($\eta_s = 10^{-5}$, $\eta_l = 10^{-4}$) and tigher thresholds ($\eta_s = 10^{-6}$, $\eta_l = 10^{-5}$), are shown in blue and orange, respectively.}
     \label{fig:err}
  \end{minipage}
  \hfill
  \begin{minipage}[t]{0.48\textwidth}
     \centering
     \includegraphics[width=0.94\columnwidth]{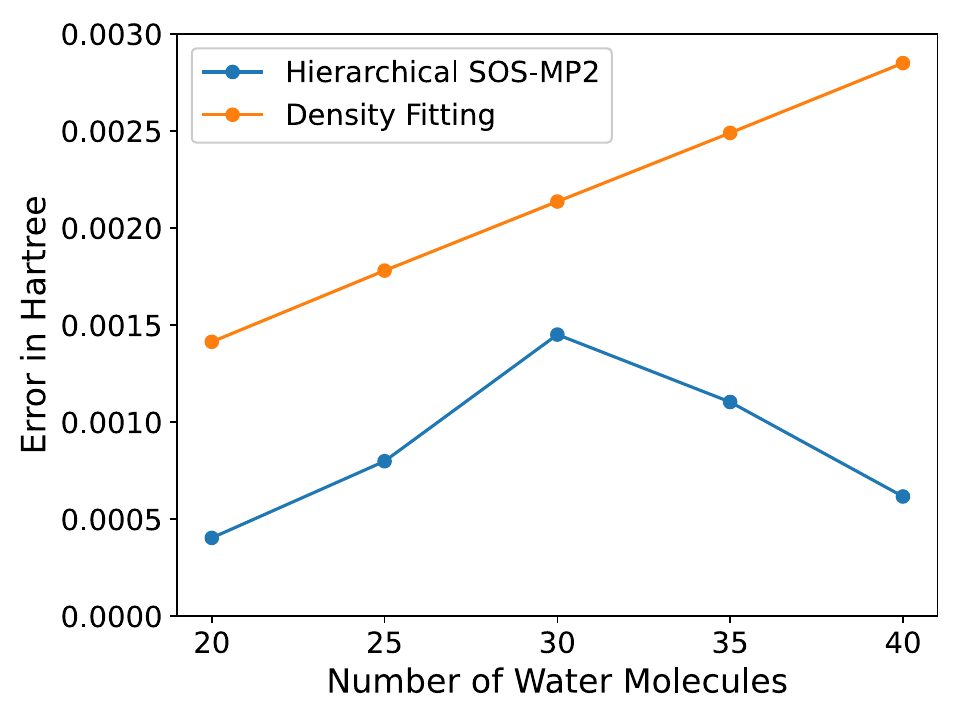}
     \caption{The error for water cluster using default thresholds.}
     \label{fig:errwat}
  \end{minipage}
\end{figure}

\subsection{Energetic Accuracy}
\label{subsec:Error analysis}

We have carried out error analysis only on small systems. To validate our method, we compute the Coulomb-like component of the MP2 energy using \textsc{Molpro}.\cite{werner2012molpro,werner2020molpro,werner2003fast}  These reference calculations were performed both exactly and with density fitting with default parameters and compared to the results from our hierarchical SOS-MP2 algorithm. Total error in energy relative to exact MP2 as a function of system size is presented for our alkane and water cluster models in Figures~\ref{fig:err} and \ref{fig:errwat}.  The observed error relative to exact MP2 is comparable to that of density fitting (DF), but achieved at significantly lower cost in both time and memory. Interestingly, the error in our results is not observed to grow linearly with system size. The dominant source of error in our implementation is due to thresholding of the $X$ and $Y$ matrices. This error behaves more like numerical rounding error and typically scales as $\mathcal{O}(\sqrt{N})$. Although this introduces some mild and irregular oscillations in the total error, it nonetheless grows more slowly than linearly.

The thresholds used for sparsification of $X$ and $Y$ can be adjusted to improve accuracy. For the time and space complexity studies presented in Sections~\ref{subsec:Space complexity analysis} and~\ref{subsec:Time complexity analysis}, we used the default thresholds listed above, which are already sufficiently small for most practical applications. However, as shown in Figure~\ref{fig:err}, this error can be reduced further by tightening the threshold by a factor of ten ($\eta_s = 10^{-6}$, $\eta_l = 10^{-5}$). Varying the threshold does not affect the qualitative trend in the sparsity growth of the $X$ and $Y$ matrices; instead, it simply shifts the onset of convergence.

For water cluster systems, we performed additional tests on smaller examples and found that the error remained quite small in practice. As illustrated in Figure~\ref{fig:errwat}, the error is notably smaller than that of DF---even though we employed relatively loose truncation thresholds for the $X$ and $Y$ matrices. We attribute this improved accuracy to cancellation effects within the sparsified matrices. Although each individual truncation introduces error, these errors tend to cancel out when summed over the full ERI contraction, resulting in significantly lower total error.

Our experiments across multiple systems and threshold values reveal a sharp drop in error near $\eta_s=1 \times 10^{-4}$~Hartree. Beyond that point, the impact of the threshold on the error slows considerably. By choosing a sparsification threshold of $\zeta = 1 \times 10^{-8}$~Hartree for the short-range part of the ERI tensor, and $\eta_s = 1 \times 10^{-8}$ for the $X$ and $Y$ matrices, the total error can be pushed below $1 \times 10^{-5}$~Hartree—comparable to the inherent numerical error of the Laplace transformation itself. However, in most cases, an error of $1 \times 10^{-4}$~Hartree (roughly 0.06 kcal/mol) is entirely acceptable, especially given that this is far smaller than the intrinsic error of MP2 itself. Accordingly, we opted for a balanced choice of thresholds (detailed in Section~\ref{subsec:The sparsity of density and density complementary matrices}) when plotting time and memory costs).

While these empirical findings are promising, we currently lack a complete theoretical understanding of the underlying cancellation behavior. It is possible that some deeper symmetry in the structure of the ERI tensor and the $X$ and $Y$ matrices drives this error suppression. We intend to explore this question further, as it may reveal new opportunities for improved performance and accuracy.

\section{Conclusion}
\label{sec:Conclusion}

The hierarchical SOS-MP2 algorithm employs a hierarchically approximated ERI tensor to accelerate the computation of SOS-MP2 correlation energies. The method is designed to leverage the sparsity of the energy-weighted density and complementary matrices, $X$ and $Y$.  A theoretical time and space complexity of $\mathcal{O}(N^2 \log N)$ is demonstrated.  The algorithm is also highly parallelizable and, in principle, could attain constant-time performance in the limit of infinite computational resources.

Our numerical experiments on both alkane chains and water clusters indicate that the measured time and space complexities scales better than $\mathcal{O}(N^2)$ in practice for larger systems, somewhat better than theoretical expectations. The performance is strongly governed by the sparsity of the $X$ and $Y$ matrices. In terms of accuracy, the error introduced by the hierarchical SOS-MP2 method is comparable to that of density fitting, but grows sub-linearly with system size and exhibits mild oscillations. This makes the method a viable and cost-effective alternative to conventional SOS-MP2 implementations, offering a favorable tradeoff between accuracy and efficiency.

There remain several opportunities for improvement. One direction involves optimizing the partitioning strategy to strike a better balance between block size and the number of hierarchical levels. At present, these parameters are tuned manually, but a more principled optimization scheme is under development. In parallel, we are pursuing a GPU-accelerated version of the algorithm, which will be tested on larger systems.

In addition to continued implementation work, we aim to better understand the cancellations observed during index transformation. Preliminary results suggest that these cancellations contribute significantly to the observed accuracy, yet the underlying mechanism remains unclear. Gaining a more complete picture of these interactions could yield new insights into the structure of the ERI tensor, and potentially inform new theoretical bounds on sparsity and error scaling.

\section*{Acknowledgements}
\label{sec:Acknowledgements}
This work was supported by an IACS seed grant. We thank Professor Robert Harrison for continued support of this project. We thank Kenneth Berard and Ying You for helpful discussions, especially on the usage of Molpro software for error comparison.  BGL acknowledges support from the National Science Foundation under grant CHE-1954519 and for start-up funding from Stony Brook University.

\section*{Author Declarations}
The authors declare no competing financial or personal interests.

\section*{Data Availability Statement}
The data that support the findings of this study are available from the corresponding author upon reasonable request.

\bibliography{refs}

\end{document}